\documentclass[prl,twocolumn,showpacs,preprintnumbers,amsmath,amssymb,superscriptaddress]{revtex4}

\usepackage{graphicx}
\usepackage{dcolumn}
\usepackage{color}

\allowdisplaybreaks
\newcommand{\beq}{\begin{equation}}
\newcommand{\eeq}{\end{equation}}
\newcommand{\beqa}{\begin{eqnarray}}
\newcommand{\eeqa}{\end{eqnarray}}

\begin{document}
\title{Electron-phonon interaction and antiferromagnetic correlations}

\author{G.~Sangiovanni}
\affiliation{Max-Planck Institut f\"ur Festk\"orperforschung, Heisenbergstr. 1,
D-70569 Stuttgart, Germany}

\author{O.~Gunnarsson}
\affiliation{Max-Planck Institut f\"ur Festk\"orperforschung, Heisenbergstr. 1,
D-70569 Stuttgart, Germany}

\author{E.~Koch}
\affiliation{Institut f\"ur Festk\"orperforschung, Forschungszentrum J\"ulich,
52425 J\"ulich, Germany}

\author{C.~Castellani}
\affiliation{
Dipartimento di Fisica Universit\`a di Roma "La Sapienza" piazzale Aldo Moro 5,
I-00185 Roma, Italy}

\author{M.~Capone}
\affiliation{INFM-SMC and Istituto dei Sistemi Complessi, Consiglio Nazionale delle Ricerche, Via dei Taurini 19, I-00185
Roma, Italy}
\affiliation{
Dipartimento di Fisica Universit\`a di Roma "La Sapienza" piazzale Aldo Moro 5,
I-00185 Roma, Italy}


\begin{abstract}
We study effects of the Coulomb repulsion on the electron-phonon 
interaction (EPI) in a model of cuprates at zero and finite doping.
We find that antiferromagnetic correlations strongly enhance EPI effects  
on the electron Green's function with respect to the paramagnetic
correlated system, but the net effect of the Coulomb interaction 
is a moderate suppression of the EPI. Doping 
leads to additional suppression, due to reduced antiferromagnetic 
correlations. In contrast, the Coulomb interaction strongly 
suppresses EPI effects on phonons, but the suppression weakens 
with doping. 
\end{abstract}
\date{\today}
\maketitle

There are strong experimental indications that the electron-phonon
interaction (EPI) plays a substantial role for properties of
high-$T_c$ cuprates \cite{Shen,Pint}, and that it even can lead to 
formation of small polarons for undoped cuprates \cite{Khyle}. 
The Coulomb interaction in the copper-oxide plane is expected to 
strongly suppress charge fluctuations. This is often described 
in the Hubbard or $t$-$J$ \cite{Zhang} models, for which important 
phonons couple to such charge fluctuations \cite{OlivertJ}. 
One might therefore actually expect the Coulomb interaction to 
strongly suppress the EPI. This was indeed found with dynamical 
mean-field theory \cite{DMFT,DMFTRPM} calculations in the 
paramagnetic phase (P-DMFT) \cite{Giorgio1,Giorgio2,Giorgio3}. On the 
other hand, calculations for the undoped antiferromagnetic $t$-$J$ 
model using the self-consistent Born approximation (SCBA) \cite{Ramsak}, 
or approximations going beyond the SCBA \cite{Mishchenko}, found that
the Coulomb interaction enhances EPI effects on the electron 
Green's function. 

The half-filled Hubbard model becomes an insulator for
large $U$. In the P-DMFT the only mechanism for this is 
the quasiparticle weight $Z$ going to zero \cite{DMFTRPM}. 
Such a reduction of $Z$ tends to strongly suppress the 
EPI \cite{Ramsak,Giorgio3}, which is an important reason 
for the small effects of the EPI in the P-DMFT. By
 allowing for antiferromagnetism (AF) in the DMFT 
\cite{Costi,Hofstetter,Giorgio4}, it is possible to 
have an insulating state although $Z$ remains finite. This 
suggests that it is important to allow for AF when describing 
the EPI, i.e., using an AF-DMFT. This is also suggested by 
the fact that in the SCBA, the enhancement of the EPI 
grows with the value of the exchange constant $J$ \cite{Ramsak}, 
i.e., with the importance of the AF. 

Here we therefore use an AF-DMFT formalism and we first consider 
a half-filled system. In contrast to previous work, we 
find that within the Holstein-Hubbard model the effects 
of the EPI on the electron Green's function are neither 
strongly suppressed nor enhanced by the Coulomb interaction. 
While previous work could only address \cite{Ramsak,Mishchenko} the 
half-filled case, the present formalism makes it possible to treat   
doped metallic cuprates, which are of particular interest. We show
that the EPI remains important for the electron Green's function as
long as AF is important. Due to the weakening of AF correlations as 
the system is doped, we find that doping reduces the effects of the 
EPI on the electron Green's function and it weakens the tendency to 
polaron formation, in agreement with experimental results \cite{Khyle}.

We also calculate the renormalization of the phonon frequency. For 
the undoped system, we find that the Coulomb repulsion very strongly
suppresses the renormalization. As the system is doped, however, the 
renormalization of the phonon frequency increases. This is the opposite 
behavior to what we find for the electron Green's function. The width and
softening of the phonon spectral function is often used to estimate
the strength of the EPI \cite{Allen}. These results show that 
for strongly correlated systems this approach may strongly underestimate 
the EPI \cite{Oliversum,Jong2}.  

We study the Holstein-Hubbard model 
\begin{eqnarray}\label{eq:1}
H&=&
-t\sum_{\langle ij\rangle \sigma}(c^{\dagger}_{i\sigma}
c^{\phantom \dagger}_{j\sigma}+{\rm H.c.}) 
+U\sum_{i}n_{i\uparrow}n_{i\downarrow}  \\ 
&+&\omega_0\sum_ib_i^{\dagger}b_i^{\phantom \dagger}+
g\sum_{i}( n_i-1)(b_i^{\phantom \dagger}+ b_i^{\dagger}),\nonumber 
\end{eqnarray}
where $c^{\dagger}_{i\sigma}$ creates an electron with spin $\sigma$
on site $i$ and $b_i$ a phonon on site $i$, $t>0$ is a hopping integral, 
$U$ the Coulomb interaction between two electrons on the same site, 
$\omega_0$ is the phonon energy,    $g$ a coupling constant and 
$n_i$ measures the number of electrons on site $i$. We assume an
infinite-coordination Bethe lattice with the half band width $D$ 
and the density of states (DOS) $N(\varepsilon)=(2/\pi D^2)
\sqrt{D^2-\varepsilon^2}$.  We define a dimensionless coupling 
constant $\lambda=g^2/(\omega_0 D)$. For large $U$ the Hubbard 
model is approximately equivalent to the $t$-$J$ model.  For a 
two-dimensional (2d) lattice, these models are related via $J/t=D/U$.

We solve the DMFT equations for the temperature $T=0$. The 
associated impurity problem is solved using the Lanczos method. 
The Hilbert space is limited by only allowing up to $N_{ph}$ 
phonons, where $N_{ph}\sim 30$ depends on the parameters. 
The energies of and couplings to the bath levels are determined 
from a continued fraction expansion \cite{Si} for the large $U$ 
half-filled case and otherwise by a fit of the cavity Green's 
function on the imaginary axis \cite{Caffarel}. We use up to 25 
bath levels. 

\begin{figure}[ht]
\begin{center}
{\rotatebox{-90}{\resizebox{6.0cm}{!}{\includegraphics {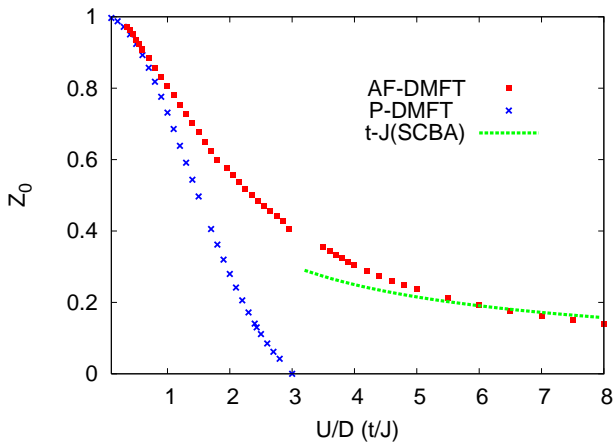}}}}
\end{center}
\caption{Quasiparticle weight $Z_0$ for $\lambda=0$ for the 
Hubbard model (as a function of $U/D$) according to P-DMFT 
and AF-DMFT and for the 2d $t$-$J$ model (as a function of 
$J/t$) according to the SCBA. The figure illustrates 
that the AF-DMFT gives reasonable values for $Z_0$.}
\label{fig1}
\end{figure}

We here focus on the quasi-particle weight $Z$, since 
our criterion for polaron formation is $Z$ being exponentially 
small and since $Z_0$, calculated for $\lambda=0$,
is expected to be crucial for the EPI, as discussed above. 
Fig.~\ref{fig1} shows $Z_0$ as a function of $U$ for the 
half-filled Hubbard model according to AF-DMFT and P-DMFT 
and as a function of $J$ for the (2d) $t$-$J$ model according to 
the SCBA. The SCBA results agree well with exact diagonalization 
results for small clusters \cite{Dagotto}. The half-filled 
system is an insulator in the AF-DMFT and for $U \gtrsim 3D$ 
in the P-DMFT \cite{Giorgio4}. While $Z_0$ drops to 
zero very quickly with $U$ in the P-DMFT, it remains finite in 
the AF-DMFT. For $U/D$ values where the Hubbard and $t$-$J$ 
models are approximately equivalent, AF-DMFT and SCBA 
agree rather well. This good description of $Z_0$ suggests that 
AF-DMFT may describe the EPI well. 

\begin{figure}[ht]
\begin{center}
\includegraphics[width=9cm]{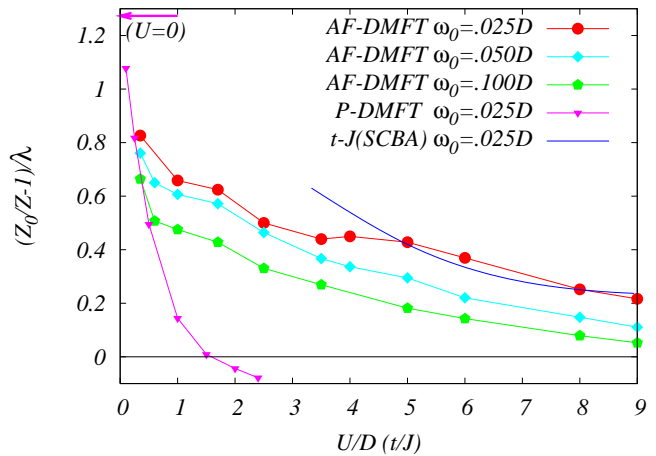}
\end{center}
\caption{$(Z_0/Z-1)/\lambda$ in the limit $\lambda \to 0$ for the 
Hubbard model according to the P-DMFT and AF-DMFT and for the 
$t$-$J$ model according to the SCBA. This quantity indicates the 
effectiveness of the EPI in the weak-coupling limit.
The figure illustrates how the EPI is much more efficient in the 
AF-DMFT than the P-DMFT, and that AF-DMFT and SCBA agree rather 
well.} 
\label{fig2}
\end{figure}

We first discuss the results for a weak EPI. For noninteracting 
electrons ($U=0$), the reduction of $Z$ by the EPI is given by 
$(1/Z-1)/ \lambda=4/\pi$ for small $\lambda$ and $\omega_0 \ll D$, 
as shown by the arrow in Fig. \ref{fig2}. To determine the effect 
of the EPI for interacting electrons ($U>0$), we calculate 
$(Z_0/Z-1)/\lambda$, shown in Fig.~\ref{fig2}. This quantity
measures how efficiently the EPI reduces the quasiparticle 
weight $Z$ with respect to $Z_0$ obtained in the absence of 
EPI. In a P-DMFT calculation it was found that the EPI very 
quickly becomes inefficient when $U$ is increased \cite{Giorgio3}. 
In contrast, allowing for AF, we find that the EPI remains much 
stronger as $U$ is increased, although it is still  
reduced compared to the noninteracting case. These results show 
that AF is crucial for the EPI of the half-filled system.  
For $U/D$ values where the Hubbard and $t$-$J$
models can be compared, the AF-DMFT and the SCBA agree well for 
$\omega_0=0.025D$. For larger phonon frequencies, however, we find  
that the EPI is appreciably more efficient in the SCBA    
than in the AF-DMFT. 

\begin{figure}[ht]
\begin{center}
\includegraphics[width=9cm]{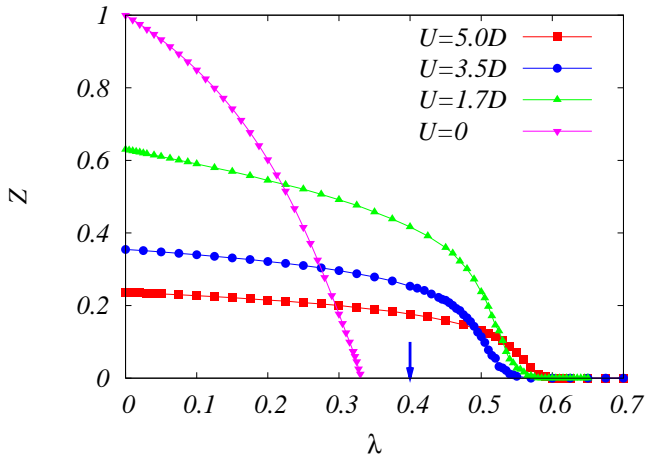}
\end{center}
\caption{$Z$ as a function of $\lambda$ for different $U$ and for 
$\omega_0=0.025D$. The arrow shows $\lambda_c$ of the $t$-$J$ 
model for $J/t=0.3$ ($U/D=3.3$) \cite{Mishchenko}. The figure 
shows how the Coulomb interaction moderately suppresses 
polaron formation ($Z \to 0$).}
\label{fig3}
\end{figure}

We next focus on strong EPI. Fig.~\ref{fig3} shows $Z$ as 
a function of $\lambda$ for different $U$. The result for $U=3.5D$ 
($J/t=0.29$) can be compared with a  calculation for the 2d $t$-$J$ 
model \cite{Mishchenko} ($J/t=0.3$ and the same $\omega_0$). As 
$\lambda$ is increased, $Z$ is strongly reduced, signaling polaron 
formation. This happens at a somewhat larger critical value 
$\lambda_c$ than was found for the $t$-$J$ model \cite{Mishchenko}, 
indicated by an arrow. The deviation from Ref. \cite{Mishchenko} 
is probably mainly due to our use of the AF-DMFT and the neglect 
of  ``crossing'' diagrams in Ref. \cite{Mishchenko}.  Good agreement 
is also found with results for the infinite dimension $t$-$J$ 
model \cite{Cappelluti}. These comparisons suggest that the AF-DMFT 
is rather accurate for the half-filled Holstein-Hubbard model.

For $U=0$, $Z$ drops very quickly as a function of $\lambda$ and 
(bi-)polarons are formed at $\lambda_c \approx 0.33$. As $U$ is 
increased the drop is slightly less rapid and polaron formation 
happens at somewhat larger $\lambda_c$. The Coulomb interaction therefore 
moderately suppresses polaron formation.

In P-DMFT calculations it was found that the effective mass 
$m^{\ast}$ depends only weakly on $\omega_0$ \cite{Giorgio3}.
In AF-DMFT we find a sizable isotope effect on $m^{\ast}$
and for $\lambda \sim \lambda_c$ the effect is  
comparable to the experimental value \cite{Khasanov}.         

\begin{figure}[ht]
\begin{center}
\includegraphics[width=9cm]{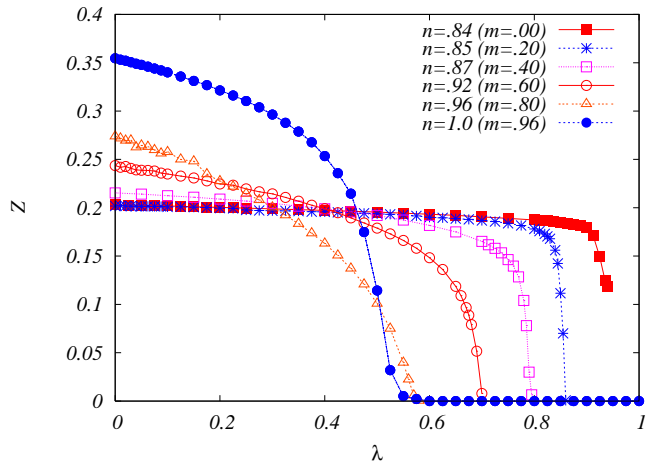}
\end{center}
\caption{$Z$ as a function of $\lambda$ for different fillings $n$ and 
associated magnetic moments $m$ for $U=3.5D$ and $\omega_0=0.025D$.
The figure illustrates how the critical $\lambda_c$ is increased 
as the filling is reduced (doping is increased) due to a reduction 
of the the antiferromagnetic correlations.}
\label{fig4}
\end{figure}

Above we have discussed the half-filled system extensively, since 
we can compare with other methods and test the reliability of the 
AF-DMFT. The doped cuprates, however, are more interesting and 
challenging, and we now focus on them. Fig.~\ref{fig4} shows $Z$ 
as a function of $\lambda$ for $U=3.5D$ and for different dopings. 
As the filling is reduced (hole doping increased) the staggered 
magnetization $m$ is reduced. The figure shows how this leads to an 
increase in $\lambda_c$. In a P-DMFT calculation \cite{Giorgio3},
on the other hand, a reduction of the filling leads to a reduction 
of $\lambda_c$. The increase of $\lambda_c$ in the AF-DMFT with 
increased doping is therefore indeed due to the reduction of $m$, 
since at constant $m=0$, $\lambda_c$ decreases with doping.

We find the AF-P transition for $U=3.5D$ at $n=0.84$,      
corresponding to a much larger doping (0.16) than found 
experimentally. This is only partly due to our neglect of second 
nearest neighbor hopping which would introduce magnetic frustration
in the system. The main reason is that in the P state there
are also AF correlations which lower the energy, but which are 
neglected in a DMFT calculation. The AF-DMFT calculation 
therefore favors the AF state. To obtained a balanced 
treatment it is necessary to use a cluster DMFT method 
\cite{Jarrell,Kotliar}. Such a calculation would introduce 
AF correlations also in the P state, and like in the AF-DMFT
calculation these correlations would weaken as the doping is 
increased. This should increase $\lambda_c$ with doping in a 
qualitative similar way as in Fig.~\ref{fig4}.

Experimentally, polaron formation is found to disappear
as the system is moderately doped \cite{Khyle}. This may be
partly due to screening of the EPI, leading to a reduction 
of $\lambda$. However, the suppression of polaron formation with 
doping for fixed $\lambda$, illustrated in Fig.~\ref{fig4}, 
should also be an essential part of the explanation. 

To study effects of the Coulomb interaction, earlier
work compared with a Holstein model with a single 
electron at the bottom of the band \cite{Ramsak,Mishchenko}. 
A better comparison, however, is with a half-filled 
Holstein model, since we can then increase $U$ keeping 
the number of electrons unchanged. For a half-filled 2d 
Holstein model such a comparison is difficult, due  to Fermi 
surface nesting and phase transitions. In DMFT calculations 
using a Bethe lattice this is not a problem. 

The EPI appears much stronger for the half-filled Holstein 
model than for a single electron at the bottom of the band. 
Comparing with the half-filled case, we therefore find that 
the EPI is suppressed by the Coulomb interaction, while earlier 
studies, comparing with the single electron case, found 
an enhancement. 

To discuss the difference between the two cases for a weak EPI, 
we calculate the electron-phonon part of the electron 
self-energy ($g$ and $\omega$ small, $\omega_0\ll D$)
\begin{equation}\label{eq:linear}
{\rm Re}\Sigma_{\rm ep}(\omega)=-\alpha {2g^2\over 
8\pi \omega_0 t}\omega,
\end{equation}
where $\alpha=1$ for the 2d single electron case but 
$\alpha=4$ for the half-filled case with a semi-elliptical
DOS. This large difference is partly due to  the DOS of 
the 2d Holstein model being smallest at the bottom of the 
band, and partly due to Re $\Sigma_{\rm ep}$ having 
contributions both from higher and lower states for 
the half-filled case, but only from higher states for the 
single electron case.                       

To understand the difference for a strong EPI, we study polaron 
formation in the adiabatic limit by comparing states  with free 
electrons and perfectly localized electrons \cite{Capone}. We find 
$E_{\rm free} =-4\beta t$ per electron, where $\beta=1$ (one electron) or  
$\beta=4/(3\pi) \approx 0.42$ (half-filled case), and $E_{\rm loc}
=-g^2/ \omega_ {\rm ph}$ per electron for both cases. We assume that 
polarons form when $|E_{\rm loc}|> |E_{\rm free}|$. This leads to 
a large $\lambda_c=1$ \cite{one} for a single electron and a much 
smaller $\lambda_c=0.42$ for the half-filled case \cite{half}.

We emphasize the remarkably small value, $\lambda_c=0.33$, for 
polaron formation in a half-filled Holstein model, meaning that 
Migdal's theorem breaks down for quite small $\lambda$. Using 
different values for $\omega_0$, small $\lambda_c$ have also 
been obtained earlier \cite{Hewson,Jong,Capone2003,Capone2005} 
using DMFT calculations.

\begin{figure}[ht]
\begin{center}
\includegraphics[width=9cm]{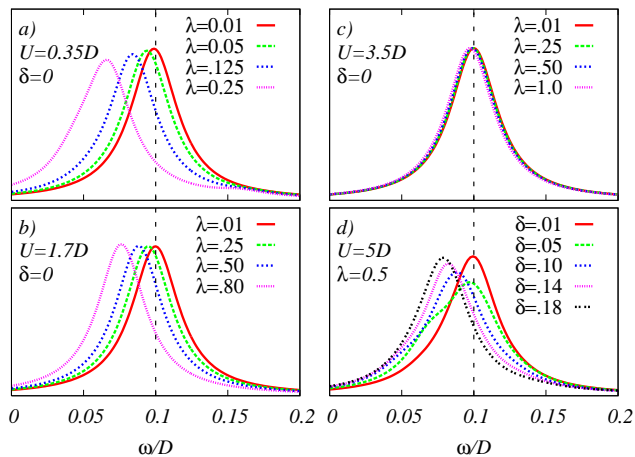}
\end{center}
\caption{Phonon spectral function for different values of 
$\lambda$. The bare phonon frequency is $\omega_0=0.1D$
and a Lorentzian broadening with the full width half
maximum of 0.04 D has been introduced. 
The figures a-c show how the phonon softening 
at half-filling is dramatically suppressed by $U$ and figure 
d that the softening increase with doping $\delta$. 
}
\label{fig5}
\end{figure}

We now consider the influence of the EPI on 
phonons. Fig.~\ref{fig5}a shows the phonon spectral function 
for a small $U$ at half-filling (doping $\delta=0$) for different
EPI strengths. The figure illustrates how the phonon is softened 
substantially as the EPI is increased. Fig.~\ref{fig5}b-c show 
this softening is strongly reduced when $U$ is increased. The 
reason is that for large $U$ charge fluctuations are strongly 
suppressed, and the system can only respond weakly to a phonon
which couples to the net charge on the atoms. Fig.~\ref{fig5}d shows 
how the softening increases as the doping is increased, due to the 
doped holes responding to phonons. This is in agreement with neutron
scattering measurements \cite{Pint}. The figure illustrates that the 
influence of the EPI on the phonon self-energy is dramatically 
different from the influence on the electron self-energy. The reason
is that the electron self-energy measures the response of the system 
to the removal or addition of a charge, which leads to a strong response
even in cases where charge fluctuations are otherwise suppressed. 

While paramagnetic DMFT calculations for the Holstein-Hubbard 
model show that effects of the EPI on {\em electrons} 
(quasiparticle weight) are very strongly suppressed by the 
Coulomb interaction, we find that this suppression is only 
moderate when antiferromagnetic (AF) correlations are  
included. As the doping is increased, the AF is reduced  
and the EPI is more suppressed. In contrast, at 
half-filling, the Coulomb interaction strongly suppresses 
effects of the EPI on  {\em phonons} (phonon softening), 
while the suppression is reduced with doping. These trends are 
consistent with experiment. 

We acknowledge useful discussions with S. Ciuchi and 
financial support (CC and MC) of Italian MIUR through Cofin 
2005 program. Calculations were performed on the J\"ulich 
JUMP computer undr Grant No.\ JIFF22.


\begin{thebibliography}{99}
\bibitem{one}This is in fairly good agreement with $\lambda_c=1.2$
from a diagrammatic Monte-Carlo calculation ($\omega_0=0.025D$)
\cite{Mishchenko} or $\lambda_c=0.84$ from a DMFT calculation
($\omega_0=0$) \cite{Ciuchi}.
\bibitem{half}This is rather close to $\lambda_c =0.33$
for $U=0$ in Fig. \ref{fig3}.
\end{thebibliography}
\end{document}